\documentclass[twocolumn,showpacs,preprintnumbers,amsmath,amssymb]{revtex4} 
\usepackage{epsfig} 
\usepackage{calc}
 
\usepackage{graphicx}
\usepackage{dcolumn}
\usepackage{bm}

\usepackage{amsthm} 
 
\newcommand{\ket}[1]{\mbox{$ | #1 \rangle $}} 
\newcommand{\bra}[1]{\mbox{$ \langle #1 | $}}

\begin{document} 
 
\title{One-way quantum key distribution: Simple upper bound on the
  secret key rate}  
 
\author{Tobias \surname{Moroder}$^{1,2}$}
\author{Marcos \surname{Curty}$^{1}$}
\author{Norbert \surname{L\"{u}tkenhaus}$^{1,2}$}
\affiliation{$^{1}$ Institute of Theoretical Physics I and Max-Planck
  Research Group, Institute of Optics, Information and Photonics,
  University Erlangen-Nuremberg, Staudtstra{\ss}e 7, 91058
  Erlangen, Germany \\ 
  $^{2}$ Institute for Quantum Computing, University
  of Waterloo, 200 University Avenue West, Waterloo, Ontario N2L 3G1, Canada}

\date{\today} 
 
\begin{abstract} 
We present a simple method to obtain an upper bound on the achievable secret
key rate in quantum key distribution (QKD) protocols that use only
unidirectional classical communication during the public-discussion
phase. This method is based on a necessary precondition for one-way secret key
distillation; the legitimate users need to prove that there
exists no quantum state having a symmetric extension that is compatible with
the available measurements results. The main advantage of the obtained upper
bound is that it can be formulated as a semidefinite program, which can
be efficiently solved. We illustrate our results by analyzing two
well-known qubit-based QKD protocols: the four-state protocol and the
six-state protocol. Recent results by Renner \textit{et al.}
~\cite{renner05a} also show that the given precondition is only
necessary but not sufficient for unidirectional secret key distillation.
\end{abstract} 
 

\maketitle 

\section{Introduction}
\label{section_introduction}

Quantum key distribution (QKD) \cite{wiesner83a, bennett84a} allows two
parties (Alice and Bob) to generate a secret key despite the computational and
technological power of an eavesdropper (Eve) who interferes with the
signals. This secret key is the essential ingredient of the one-time-pad or
Vernam cipher \cite{vernam26a}, which can provide
information-theoretic secure communications.   
 
Practical QKD protocols distinguish two phases in order to generate a secret
key: a quantum phase and a classical phase. In the quantum phase a
physical apparatus generates classical data for Alice and Bob
distributed according to a joint probability distribution
$p(a_i,b_j)\equiv p_{ij}$. In the classical phase, Alice and Bob try
to distill a secret key from $p_{ij}$ by means of a public discussion
over an authenticated classical channel.   
 
Two types of QKD schemes are used to create the correlated data $p_{ij}$. In
\emph{entanglement based} (EB) schemes, a source, which is assumed to be under
Eve's control, produces a bipartite quantum state $\rho_{AB}$ that is
distributed to Alice and Bob. Eve could even have a third system entangled
with those given to the legitimate users. Alice and Bob measure each incoming
signal by means of two \emph{positive operator valued measures} (POVM)
\cite{helstrom76a} $\{A_i\}$ and $\{B_j\}$, respectively, and they obtain
$p_{ij}=Tr(A_i\otimes{}B_j\ \rho_{AB})$. 
 
In an ideal \emph{prepare and measure} (PM) schemes, Alice prepares a
pure state $\ket{\varphi_i}$ with probability $p_i$ and sends it to
Bob. On the receiving side, Bob measures each received signal with a
POVM $\{B_j\}$. The signal 
preparation process in PM schemes can be also thought of as follows
\cite{bennett92c}: First, Alice produces the bipartite state
$\ket{\psi_{source}}_{AB}=\sum_i \sqrt{p_i} \ket{\alpha_i}_A\ket{\varphi_i}_B$
and, afterwards, she measures the first subsystem in the orthogonal basis
$\ket{\alpha_i}_A$ corresponding to the measurements
$A_i=\ket{\alpha_i}_A\bra{\alpha_i}$. This action generates the
(non-orthogonal) signal states 
$\ket{\varphi_i}$ with probabilities $p_i$. In PM schemes the reduced density
matrix of Alice, $\rho_A=Tr_B(\ket{\psi_{source}}_{AB}\bra{\psi_{source}})$,
is fixed and cannot be modified by Eve. To include this information in the
measurement process one can add to the observables $\{A_i \otimes B_j
\}$, measured by Alice and Bob, other observables $\{C_k
\otimes\openone\}$ such that they form a tomographic 
complete set of Alice's Hilbert space \cite{curty04a,curty05a}. In the
most general PM scheme Alice is free to prepare arbitrary states
$\rho_i$ instead of only pure states $\ket{\varphi_i}$. One can apply
the same framework as for the ideal PM scheme, as reviewed in
App.~\ref{ap_mixedstates}.  
 
From now on, we will consider that $p_{ij}$ and $\{A_i\otimes B_j\}$ refer
always to the complete set of measurements, \textit{i.e.}, they include also
the observables $\{C_k \otimes \openone\}$ for PM schemes.   
 
The public discussion performed by Alice and Bob during the classical phase of
QKD can involve either one-way or two-way classical communication. Two-way
classical communication is more robust than one-way in terms
of the amount of errors that the QKD protocol can tolerate in order to distill
a secret key \cite{gottesman03a}. However, the first security proof of
QKD by Mayers \cite{mayers96a}, and the most commonly known proof by
Shor and Presskill \cite{shor00a} are based on one-way communications,
and many other security proofs of QKD belong also to this last paradigm
\cite{lo01a,tamaki03a}. Moreover, any two-way communication protocol
includes a final non-trivial step that is necessarily only one-way, so
that the study of one-way communication is also useful for the study
of two-way communication.  

In this paper we concentrate on one-way classical communication 
protocols during
the public discussion phase. Typically, these schemes consist of three
steps: local pre-processing of the 
data, information reconciliation to correct the data, and privacy
amplification to decouple the data from Eve \cite{luetkenhaus99a}. Depending
on the allowed direction of communication, two different cases must be
considered. \emph{Direct communication} refers to communication from Alice to
Bob, \emph{reverse reconciliation} allows communication from Bob to Alice
only. (See, for instance, \cite{gross03a,heid05a}.) We will consider
only the case of direct communication. Expressions for the opposite
scenario, \textit{i.e.}, reverse reconciliation, can be directly
obtained simply by renaming Alice and Bob. Note that for typical
experiments, the joint probability distribution $p_{ij}$ is not
symmetric, so that the qualitative statements for both cases will
differ. 

We address the question of how much secret key can be
obtained from the knowledge of $p_{ij}$ and $\{A_i\otimes B_j\}$. This
is one of the most important figures of merit in order to compare the
performance of different QKD schemes.  We consider the so-called \emph{trusted
device scenario}, where Eve cannot modify the actual detection devices 
employed by Alice and Bob. (See Refs. \cite{curty05a,curty04b}.) We assume
that the legitimate users have complete knowledge about their detection
devices, which are fixed by the actual experiment.    

In the last years, several lower and upper bounds on the secret key
rate for particular one-way QKD schemes have been proposed. The lower
bounds come from protocols that have been proven to be secure
\cite{shor00a,lo01a,tamaki03a,devetak03a,kraus04a,renner05a}. The
upper bounds are generally derived by considering some particular
eavesdropping attack and by determining when this attack can defeat
QKD
\cite{fuchs97a,cirac97a,bech99a,kraus04a,renner05a}. Unfortunately, to
evaluate these known bounds for general QKD protocols is not always a
trivial task. Typically, it demands to solve difficult optimization
problems, which can be done only for some particular QKD protocols
\cite{renner05a}.   
 
In this paper we present a simple method to obtain an upper bound on the
secret key rate for general one-way QKD protocols. 
The obtained upper bound will not be tight
for all QKD schemes, but it has the advantage that it is straightforward to
evaluate in general since it can be formulated as a semidefinite program
\cite{vandenberghe96a,vandenberghe04a}. Such instances of convex optimization
problems can be efficiently solved, for example by means of interior-point
methods \cite{vandenberghe96a,vandenberghe04a}. Our analysis is based on a
necessary precondition for one-way QKD: The legitimate
users need to prove that there exists no quantum state having a symmetric
extension that is compatible with the
available measurement results \cite{symmetric_extension}. 
This kind of states (with symmetric extensions)
have been recently analyzed in
Refs. \cite{doherty02a,doherty04a,doherty05a}. 

The paper is organized as follows. In Sec.~\ref{section_knownbounds} we review
some known upper bounds on the secret key rate using one-way post-processing
techniques. Sec.~\ref{section_newbound} includes the main result of the paper.
Here we introduce a straightforward method to obtain an upper bound on
the secret key rate for one-way QKD. This result is then illustrated in
Sec.~\ref{section_evaluation} for two well-known qubit-based QKD protocols:
the four-state \cite{bennett84a} and the six-state \cite{bruss98a} QKD
schemes. We select these two particular QKD schemes because they allow us to
compare our results with already known upper bounds in the literature
\cite{fuchs97a,cirac97a,bech99a,kraus04a,renner05a}. Then in 
Sec.~\ref{section_conclusion} we present our conclusions. The paper includes 
also two Appendices. In
App.~\ref{ap_mixedstates} we consider very briefly the case of 
QKD based on mixed signal states instead of pure states. Finally, 
App.~\ref{ap_sdpmethod} contains the semidefinite
program needed to actually solve the upper bound derived in 
Sec.~\ref{section_newbound}.

\section{Known Upper Bounds} 
\label{section_knownbounds}
 
Different upper bounds on the secret key rate for one-way QKD have been
proposed in the last years. These results either apply to a specific QKD
protocol \cite{fuchs97a,cirac97a,bech99a}, or they are derived for different
starting scenarios of the QKD scheme
\cite{devetak03a,kraus04a,renner05a}, \textit{e.g.}, one where Alice
and Bob are still free to design suitable measurements. 

Once Alice and Bob have performed their measurements during the quantum phase
of the protocol, they are left with two classical random variables $A$ and $B$,
respectively, satisfying an observed joint probability distribution
$p(a_i,b_j)\equiv p_{ij}$. On the other hand, Eve can keep her quantum
state untouched and delay her measurement until the public-discussion
phase, realized by Alice and Bob, has finished.  

In order to provide an upper bound on the secret key rate it is sufficient 
to consider a particular eavesdropping strategy. For instance, we can restrict
ourselves to collective attacks \cite{kraus04a,renner05a}. 
This situation can be modelled by assuming
that Alice, Bob, and Eve share an unlimited number of the so-called ccq states
$\rho_{ccq}$ which are given by \cite{devetak03a}  
\begin{equation}
  \label{nuevo}
  \rho_{ccq} = \sum_{i,j}   p_{ij} \ket{ij}_{AB}\bra{ij} \otimes
  \rho^{i,j}_{E}, 
\end{equation}
where $\rho^{i,j}_E$ denotes Eve's conditional quantum state, and the states
$\{\ket{i}_{A} \}$ and $\{ \ket{j}_{B} \}$ form orthonormal basis sets  
for Alice and Bob, respectively. As shown in Refs.~\cite{kraus04a,renner05a},
in  this scenario the rate $K_\rightarrow$, at which Alice and Bob
can generate a secret key by using only direct communication, is bounded from
above by  
\begin{equation} 
  \label{upperbound_ccq} 
  K_\rightarrow \leq \mathop{\sup_{\sigma_U \leftarrow A}}_{\sigma_T
  \leftarrow A}  S(U|ET) - S(U|BT),   
\end{equation} 
where the supremum is taken over all possible density operators $\sigma_U$ and
$\sigma_T$ depending on the random variable $A$ of Alice. The von
Neumann entropy of a quantum state $\rho$ reads as $S(\rho)=-Tr(\rho
\log \rho)$, while the conditional von Neumann entropy $S$ is defined
in terms of von Neumann entropies itself, \textit{i.e.},
$S(U|ET)=S(UET)-S(ET)$. The upper bound given by
Eq.~(\ref{upperbound_ccq}) refers to the quantum state given by
Eq.~(\ref{nuevo}) after a local post-processing step. It is given by
\cite{kraus04a,renner05a} 
\begin{equation}
  \rho_{UTBE}=\sum_{i,j} p_{ij}\ \sigma_U^i \otimes \sigma_T^i \otimes
  \ket{j}_B\bra{j} \otimes \rho^{i}_{E},
\end{equation}
where $\rho_E^i=\sum_j p(b_j|a_i) \rho_E^{i,j}$. 
This upper
bound involves only a single letter optimization problem. However, 
the optimization runs over density operators $\sigma_U$ and $\sigma_T$ which 
makes Eq.~(\ref{upperbound_ccq}) hard to evaluate. 
 
Another upper bound that applies to the QKD scenario that we consider here is
the Csisz\'ar and K\"orner's secret key rate for the one-way \emph{classical}
key-agreement scenario \cite{csiszar78a}. Suppose that Alice, Bob, and Eve have
access to many independent realizations of three random variables $A$, $B$,
and $E$, respectively, that are distributed according to the joint probability
distribution $p(a_i,b_j,e_k)$. Csisz\'ar and K\"orner showed that the one-way
secret key rate is given by \cite{csiszar78a}
\begin{equation} 
  \label{rate_classicalbound}
  S_\rightarrow(A;B|E)= \mathop{\sup_{U \leftarrow A}}_{T \leftarrow U}
  H(U|ET) - H(U|BT).
\end{equation} 
The single letter optimization ranges over two classical channels
characterized by  
the transition probabilities $Q(u_l|a_i)$ and $R(t_m|u_l)$, and where the
conditional Shannon entropy is defined as $H(U|ET)=- \sum p(u_l,e_k,t_m) \log
p(u_l|e_k,t_m)$. The first channel produces the secret key $U$, while the
second channel creates the broadcasted information $T$. 

Note that Eq.~(\ref{rate_classicalbound}) provides also an upper bound on
$K_\rightarrow$. Eve can always measure her subsystem of the ccq state 
given by Eq.~(\ref{nuevo}) by means of a POVM $\{ E_k \}$. As a result, 
Alice, Bob, and Eve share the
tripartite probability distribution $p(a_i,b_j,e_k)=p_{ij}\ Tr(E_k
\rho_E^{i,j})$. Unfortunately, the optimization problem that one 
has to solve in order to obtain $S_\rightarrow(A;B|E)$ is also non-trivial, 
and its solution is only known for particular examples. 
(See Ref.~\cite{holenstein05a}.)     
 
Finally, an easy computable upper bound on $K_\rightarrow$ is given by the
classical 
mutual information $I(A;B)$ between Alice and Bob \cite{maurer99a}. This
quantity is defined in terms of the Shannon entropy $H(A)=-\sum  
p(a_i)\log{p(a_i)}$ and the Shannon joint entropy $H(A,B)=-\sum
p(a_i,b_j)\log{p(a_i,b_j)}$ as 
\begin{equation}
I(A;B) = H(A) + H(B) - H(A,B). 
\end{equation}
The mutual information represents an upper bound on the secret key rate for
\textit{arbitrary} public communication protocols, hence in particular for
one-way communication protocols \cite{maurer99a}, {\it i.e.}, 
\begin{equation}
  \label{rate_mutualinformation}
  K_\rightarrow \leq S_\rightarrow(A;B|E) \leq I(A;B).  
\end{equation}
To evaluate $I(A;B)$ for the case of QKD, we only need to use as $p(a_i,b_j)$
the correlated data $p_{ij}$.

\section{Upper Bound on $K_\rightarrow$} 
\label{section_newbound}
 
Our starting point is again the observed joint probability distribution
$p_{ij}$ obtained by Alice and Bob after their measurements. This probability
distribution defines an equivalence class $\mathcal{S}$ of quantum states  
that are compatible with it,  
\begin{equation}
  \label{eq_class} 
  \mathcal{S}=\left\{ \rho_{AB}\ |\ Tr(A_i \otimes B_j\  
  \rho_{AB})=p_{ij}, \ \forall i,j \right\}.
\end{equation} 
By definition, every state $\rho_{AB} \in
\mathcal{S}$ can represent the state shared by Alice and Bob before their
measurements \footnote{The equivalence class $\mathcal{S}$ reduces to the
  trivial one, \textit{i.e.}, it contains  only one element up to a global
  phase, when the measurements realized by Alice and Bob provide complete
  tomographic information about $\rho_{AB}$. This is the case, for instance,
  in the six-state QKD protocol \cite{bruss98a}. Otherwise, $\mathcal{S}$
  contains always more than one quantum state.}.

Now the idea is simple: just impose some \emph{particular} eavesdropping
strategy for Eve, and then use one of the already known upper bounds. (See
also Ref.~\cite{moroder05a}.) The upper bound resulting represents an upper
bound for \emph{any} possible eavesdropping strategy. The method can be
described with the following three steps.  

(1) Select a particular eavesdropping strategy for Eve. This strategy is given
by the choice of a tripartite quantum state $\rho_{ABE}$ and a POVM $\{E_k\}$
to measure Eve's signals. The only restriction here is that the chosen
strategy cannot alter the observed data, \textit{i.e.}, $Tr_{E}(\rho_{ABE})\in
\mathcal{S}$.  
 
(2) Calculate the joint probability distribution $p_{ijk}=Tr(A_i
\otimes B_j \otimes E_k \ \rho_{ABE})$.  
 
(3) Use an upper bound for $K_\rightarrow$ given the probability distribution 
$p_{ijk}$. Here we can use,
for instance, Eq.~(\ref{rate_classicalbound}) or just the mutual information 
between Alice and Bob which
is straightforward to calculate.  
 
This method can be improved by performing an optimization over all possible
measurements on Eve's system and over all possible tripartite states that Eve
can access \footnote{Note that this improvement still does not represent the
  most general case, since it forces Eve to measure directly her state before
  the public-discussion phase. In principle, Eve might also use eavesdropping
  strategies where no measure on her system is performed until the
  public-discussion phase is completed, or she could even wait to attack the
  final cryptosystem, that uses the generated key, with her quantum probe. This
  last scenario is related with the idea of composability
  \cite{renner04b,benor04a,benor04b}.}.   
This gives rise to a set of possible extensions $\mathcal{P}$ of the observed 
bipartite probability distribution $p_{ij}$ for the random variables $A$ and 
$B$ to a tripartite probability distribution $p_{ijk}$ for the random
variables $A$, $B$, and $E$. Now the upper bound is given by
\begin{equation}
  \label{upperbound_eavesdropping}
  K_\rightarrow \leq \inf_{\mathcal{P}} S_\rightarrow,
\end{equation}
with $S_\rightarrow$ representing the chosen quantity in step (3).

In Sec.~\ref{sec3a} we present a necessary precondition for one-way QKD. 
In particular, Alice and Bob need to prove that there exists
no quantum state having a
symmetric extension that is 
compatible with the available measurements results 
\cite{symmetric_extension}. Motivated by this necessary
precondition, we introduce a special class of eavesdropping strategies for
Eve in Sec.~\ref{section_eavesdroppingmodel}. These strategies are based on a
decomposition of quantum states similar to the best separability approximation
\cite{lewenstein97a,karnas01a}, but now for states with symmetric
extensions. The general idea followed here is similar to that
presented in Ref.~\cite{moroder05a} for two-way upper bounds on QKD. 

\subsection{States With Symmetric Extensions \& One-Way QKD}\label{sec3a} 

A quantum state $\rho_{AB}$ is said to have a symmetric extension to two
copies of system $B$ if and only if there exists a tripartite state
$\rho_{ABB^\prime}$ with $\mathcal{H}_B=\mathcal{H}_{B^\prime}$ which fulfills
the following two properties \cite{doherty02a}: 
\begin{eqnarray}
  Tr_{B^\prime}(\rho_{ABB^\prime}) & = & \rho_{AB},\\
  P \rho_{ABB^\prime} P & = & \rho_{ABB^\prime},
\end{eqnarray}
where the operator $P$ satisfies $P \ket{ijk}_{ABB^\prime} =
\ket{ikj}_{ABB^\prime}$. This definition can be easily extended to cover also
the case of symmetric extensions of $\rho_{AB}$ to two copies of system $A$,
and also of extensions of $\rho_{AB}$ to more than two copies of system $A$ or
of system $B$.   

States with symmetric extension play an important role in quantum information
theory, as noted recently. They can deliver a complete family of separability
criteria for the bipartite \cite{doherty02a, doherty04a} and for the
multipartite case \cite{doherty05a}, and they provide a constructive way to
create local hidden variable theories for quantum states
\cite{terhal03a}. Moreover, they are related to the capacity of quantum
channels \cite{horodecki05apre}. Most important, a connection to one-way QKD
has also been noticed: 

\textit{Observation 1~\cite{symmetric_extension}:} If the observed
data $p_{ij}$ originate from a quantum 
state $\rho_{AB}$ which has a symmetric extension to two copies of system $B$,
then the secret key rate for unidirectional communication
$K_\rightarrow$ from Alice to Bob vanishes.  

\textit{Proof:} Suppose that the observed data $p_{ij}$ originate from a state
$\rho_{AB}$ which has a symmetric extension to two copies of system
$B$. Suppose as well that the third subsystem of the extended tripartite state
$\rho_{ABB^\prime}$ is in Eve's hands, \textit{i.e.},
$\rho_{ABE}=\rho_{ABB^\prime}$. This results in equal marginal states for
Alice-Bob and Alice-Eve, \textit{i.e.}, $\rho_{AB}=\rho_{AE}$. From Alice's
perspective the secret key distillation task is then completely symmetric
under interchanging Bob and Eve. Since we restrict ourselves to unidirectional
classical communication from Alice to Bob only, we find that it is impossible
for Bob to break this symmetry. That is, if Alice tries to generate 
a secret key with Bob her actions would automatically create exactly the same
secret key with Eve. To complete the proof we need to verify that Eve can
access the symmetric extension $\rho_{ABB^\prime}$ of $\rho_{AB}$ in both
kinds of QKD schemes, EB schemes and PM schemes. It was demonstrated in
Ref.~\cite{curty04a} that Eve can always create a purification of the original
state $\rho_{AB}$, which means that Eve can have access to the symmetric
extension. $\blacksquare$    

\textit{Remark 1:} A quantum state $\rho_{AB}$ has a symmetric
extension to two copies of system $B$ if and only if there exists a
tripartite state $\rho_{ABE}$ with equal marginal states for Alice-Bob
and Alice-Eve, \textit{i.e.}, $\rho_{AB}=\rho_{AE}$.  

\textit{Proof:} If a quantum state $\rho_{AB}$ has a symmetric
extension this automatically implies equal marginal states for
Alice-Bob and Alice-Eve. For the other direction, suppose that there
exists a tripartite state $\tilde \rho_{ABE}$ with equal marginals,
but which is not symmetric under interchange of subsystems $B$ and
$E$. Then the state $P \tilde \rho_{ABE} P$ is also a possible
tripartite state with equal marginals. This allows to construct the
symmetric extension of the state $\rho_{AB}$ as $\rho_{ABE}=1/2(\tilde \rho_{ABE}
+ P \tilde \rho_{ABE} P)$. $\blacksquare$ 

There exists entangled states which do have symmetric
extensions \cite{doherty02a,doherty04a}. Hence, accordingly to Observation $1$,
although these states are entangled and therefore potentially useful
for two-way QKD~\cite{curty04a}, they are nevertheless useless for one-way
QKD in the corresponding direction.
 
We define the best extendibility approximation of a given state $\rho_{AB}$ as
the decomposition of $\rho_{AB}$ into a state with symmetric extension, that
we denote as  $\sigma_{ext}$, and a state without symmetric extension
$\rho_{ne}$, while maximizing the weight of the extendible part $\sigma_{ext}$ 
\footnote{From
  now on, the term extension will always stand for a symmetric extension 
  to two copies of system $A$ or $B$. We will not make any further distinction
  between the different types of extension and we simply call the state
  extendible. The extension to two copies of system $A$ corresponds to reverse
  reconciliation, and extensions to two copies of system $B$ corresponds to
  the direct communication case.}, \textit{i.e.},
\begin{equation}
  \label{bse}
  \rho_{AB}=\max_\lambda \lambda \sigma_{ext} + (1-\lambda) \rho_{ne}.
\end{equation}
This definition follows the same spirit as the best separability
approximation introduced in Refs.~\cite{lewenstein97a, karnas01a}. Since the
set of all extendible quantum states forms a closed and convex set
\cite{doherty04a}, the maximum in Eq.~(\ref{bse}) 
always exists. We denote the maximum weight of
extendibility 
of $\rho_{AB}$
as $\lambda_{max}(\rho_{AB})$, where $0 \leq
\lambda_{max}(\rho_{AB}) \leq 1$ is satisfied.  

Given an equivalence class $\mathcal{S}$ of quantum states, we define the
maximum weight of extendibility within the equivalence class, denoted as
$\lambda_{max}^{\mathcal{S}}$, as  
\begin{equation} 
  \label{lambda_max} 
  \lambda_{max}^{\mathcal{S}} = \max \left \{ \lambda_{max}(\rho_{AB})\ |\
  \rho_{AB} \in \mathcal{S} \right \}.  
\end{equation} 
This parameter is related to the necessary precondition for one-way secret key
distillation by the following observation. 

\textit{Observation 2}: Assume that Alice and Bob can perform local
measurements with POVM elements $A_i$ and $B_j$, respectively, to obtain the
joint probability distribution of the outcomes $p_{ij}$ on the distributed
quantum state $\rho_{AB}$. Then the following two statements are equivalent:
(1) The correlations $p_{ij}$ can originate from an extendible state. (2) The
maximum weight of extendibility $\lambda_{max}^{\mathcal{S}}$ within the
equivalence class of quantum states $\mathcal{S}$ compatible with the observed
data $p_{ij}$ satisfies $\lambda_{max}^{\mathcal{S}}=1$.  
 
\textit{Proof}. If $p_{ij}$ can originate from an extendible state, then there
exists a $\sigma_{ext}$ such as $\sigma_{ext} \in \mathcal{S}$. Moreover, we
have that any extendible state satisfies $\lambda_{max}(\sigma_{sep})=1$. The
other direction is trivial. $\blacksquare$  

Let us define $\mathcal{S}_{max}$ as the equivalence class of quantum states
composed of those states $\rho_{AB}\in\mathcal{S}$ that have maximum weight of
extendibility. It is given by 
\begin{equation} 
  \mathcal{S}_{max} = \left \{ \rho_{AB}\in\mathcal{S}\  |\ 
  \lambda_{max}(\rho_{AB})=\lambda_{max}^{\mathcal{S}} \right \}. 
\end{equation}

\subsection{Eavesdropping Model}
\label{section_eavesdroppingmodel}
 
An eavesdropping strategy for our purpose is completely characterized by
selecting the overall tripartite quantum state $\rho_{ABE}$ and the
measurement operators $\{ E_k \}$. Again, the only restriction here is that
$Tr_E(\rho_{ABE}) \in \mathcal{S}$. We consider that Eve chooses a purification
$\rho_{ABE}=\ket{\Phi}_{ABE}\bra{\Phi}$ of a state $\rho_{AB}$ taken from the
equivalence class $\mathcal{S}_{max}$.  

The quantum states $\sigma_{ext}$ and $\rho_{ne}$ of the best extendibility 
approximation of $\rho_{AB}$ can be written in terms of their spectral
decomposition as \footnote{If the best extendibility approximation of the 
  state $\rho_{AB}$ is not unique, Eve simply takes one particular 
  decomposition of the possible set of them.} 
\begin{eqnarray}  
  \sigma_{ext} &=& \sum_i q_i \ket{\phi_i}_{AB}\bra{\phi_i}, \\ 
  \rho_{ne} &=& \sum_i p_i \ket{\psi_i}_{AB}\bra{\psi_i}, 
\end{eqnarray} 
with $\bra{\phi_i}\phi_j\rangle=\bra{\psi_i}\psi_j\rangle=0$ for all
$i\neq{}j$. A possible purification of the state $\rho_{AB}$ is given by 
\begin{eqnarray}
  \label{purification} 
  \ket{\Phi}_{ABE}=\sum_i \sqrt{\lambda_{max}^{\mathcal{S}} q_i} 
  \ket{\phi_i}_{AB}\ket{t_k=ext,f_i}_E + \nonumber \\ 
  \sum_j \sqrt{(1-\lambda_{max}^{\mathcal{S}})p_j} 
  \ket{\psi_j}_{AB}\ket{t_k=ne,f_j}_E, 
\end{eqnarray} 
where the states $\{\ket{t_k=ext,f_i},\ket{t_k=ne,f_j}\}$ form an
orthogonal bases on Eve's subsystem.

It is important to note that in both kinds of QKD schemes, EB schemes and PM
schemes, Eve can always have access to the state $\ket{\Phi}_{ABE}$ given by
Eq. (\ref{purification}). This has been shown in
Ref.~\cite{curty04a}. In an EB scheme this is clear since Eve is the
one who prepares the state $\rho_{AB}$ and who distributes it to Alice
and Bob. In 
the case of PM schemes we need to show additionally that Eve can
obtain the state $\ket{\Phi}_{ABE}$ by interaction with Bob's system
only. In the Schmidt decomposition the state prepared by Alice,
$|\psi_{source}\rangle_{AB}$, can be written as $|\psi_{source}\rangle =
\sum_i c_i |u_i\rangle_A |v_i\rangle_B$. Then the Schmidt decomposition of
$\ket{\Phi}_{ABE}$, with respect to system $A$ and the composite system $BE$,
is of the form $\ket{\Phi}_{ABE}= \sum_i\ c_i |u_i\rangle_A
|\tilde{e}_i\rangle_{BE}$ since $c_i$ and $|u_i\rangle_A$ are fixed by the
known reduced density matrix $\rho_A$ to the corresponding values of
$|\psi_{source}\rangle_{AB}$. Then one can find a suitable unitary operator
$U_{BE}$ such that $|\tilde{e}_i\rangle_{BE}=U_{BE}|v_i\rangle_B|0\rangle_E$
where $|0\rangle_E$ is an initial state of an auxiliary system. 
 
For simplicity, we consider a special class of measurement strategies for Eve.
This class of measurements can be thought of as a two step procedure: 
 
(1) First, Eve distinguishes contributions coming from the part with symmetric
extension and from the part without symmetric extension of $\rho_{AB}$. The
corresponding measurements are projections of Eve's subsystem onto the
orthogonal subspaces $\Pi_{ext}=\sum_i \ket{t_k=ext,f_i}\bra{t_k=ext,f_i}$ and
$\Pi_{ne}=\sum_j \ket{t_k=ne,f_j}\bra{t_k=ne,f_j}$.   
 
(2) Afterwards, Eve performs a refined measurement strategy on each 
subspace separately. As we will see, only the non-extendible part
$\rho_{ne}$ might allow Alice and Bob to distill a secret key by direct
communication; from the extendible part no secret key can be obtained.  

We shall label Eve's measurement outcomes $e_k$
with two variables, $e_k=(t_k,f_k)$. The
first variable $t_k\in\{ext,ne\}$ denotes the outcome of the
projection measurement, while $f_k$ corresponds to the outcome arising
from the second step of the measurement strategy. With probability
$p(t_k=ne)=1-\lambda_{max}^\mathcal{S}$ Eve finds that Alice and Bob
share the non-extendible part of $\rho_{AB}$. After this first
measurement step, the conditional quantum state shared by Alice, Bob,
and Eve, denoted as $\rho_{ABE}^{ne} = \ket{\Phi_{ne}}
\bra{\Phi_{ne}}$, corresponds to a purification of $\rho_{ne}$,
\textit{i.e.}, 
\begin{equation} 
\ket{\Phi_{ne}}_{ABE}=\sum_j \sqrt{p_j} 
\ket{\psi_j}_{AB}\ket{t_k=ne,f_j}_E. 
\end{equation} 
 
Next we provide an upper bound for $K_{\rightarrow}$ that arises from
this special eavesdropping strategy. Moreover, as we will see, the obtained
upper bound is straightforward to calculate. 

\subsection{Resulting Upper Bound}
\label{res} 
 
For the special eavesdropping strategy considered in
Sec.~\ref{section_eavesdroppingmodel}, we will show that we can
restrict ourselves to the non-extendible part $\rho_{ne}$ of a given
$\rho_{AB}$ only. As a consequence, the resulting upper bound will
only depend on this non-extendible part. This motivates the definition
of a new equivalence 
class of quantum states $\mathcal{S}_{max}^{ne}$, defined as   
\begin{equation} 
  \mathcal{S}_{max}^{ne}= \left\{ \rho_{ne}(\rho_{AB})\ |\ \rho_{AB} \in 
  \mathcal{S}_{max} \right\},
\end{equation} 
where $\rho_{ne}(\rho_{AB})$ represents the non-extendible part in a valid
best extendibility approximation of $\rho_{AB}\in\mathcal{S}_{max}$
given by Eq. (\ref{bse}). To simplify the notation, from now on we will
write $\rho_{ne}$ instead of $\rho_{ne}(\rho_{AB})$. The possibility
to concentrate on the non-extendible parts only is given by the
following theorem.

\textit{Theorem 1}: Suppose Alice's and Bob's systems are subjected to
measurements described by the POVMs $\{ A_i \}$ and $\{ B_j \}$ respectively,
and their outcomes follow the probability distribution $p_{ij}$. They
try to distill a secret key by unidirectional classical communication
from Alice to Bob only. The secret key rate, denoted as
$K_\rightarrow$, is bounded from above by 
\begin{equation}
  \label{bound_theorem1} 
  K_{\rightarrow} \leq (1-\lambda_{max}^{\mathcal{S}}) \inf_{\mathcal{P}^*}
  S_{\rightarrow}(A;B|E),  
\end{equation} 
where $S_{\rightarrow}(A;B|E)$ denotes the classical one-way secret key rate
given by Eq.~(\ref{rate_classicalbound}) for a tripartite probability
distribution $\tilde p_{ijk} \in \mathcal{P^*}$. The set
$\mathcal{P^*}$ considers all possible POVMs $\{ E_k \}$ which Eve can
perform on a purification $\ket{\Phi_{ne}}_{ABE}$ of the
\emph{non-extendible part} $\rho_{ne} \in \mathcal{S}_{max}^{ne}$
only, \textit{i.e.}, $\tilde p_{ijk}=Tr[ A_i\otimes B_j\otimes E_k
(\ket{\Phi_{ne}}_{ABE}\bra{\Phi_{ne}})]$.  
 
\textit{Proof}: In order to derive Eq.~(\ref{bound_theorem1}) we have
considered only a particular class of eavesdropping strategies for Eve
as described in Sec.~\ref{section_eavesdroppingmodel}. This class
defines a subset $\mathcal{P}^\prime$ of the set of all possible
extensions $\mathcal{P}$ of the observed data $p_{ij}$ to a general
tripartite probability distribution $p_{ijk}$, which are considered in
the upper bound given by Eq. (\ref{upperbound_eavesdropping}). We
have, therefore, that  
\begin{equation}
  K_\rightarrow \leq \inf_{\mathcal{P}} S_\rightarrow(A;B|E) \leq
  \inf_{\mathcal{P}^\prime} S_\rightarrow(A;B|E).
\end{equation}

As introduced in Sec.~\ref{section_eavesdroppingmodel}, we label the
outcome of Eve's measurement strategy by two variables
$e_k=(f_k,t_k)$, where the value of $t_k \in \{ext, ne\}$ labels the
outcome of her projection measurement. For the tripartite
probability distribution $p(a_i,b_j,e_k=(f_k,t_k))$ we denote the
secret key rate by $S_\rightarrow(A;B|FT)$. 

For the one-way secret key rate $S_\rightarrow(A;B|FT)$ we get
\begin{eqnarray}
  \nonumber
  &&S_\rightarrow(A;B|FT)= \mathop{\sup_{U \leftarrow A}}_{V \leftarrow
    U} H(U|VFT)-H(U|VB) \\
  \nonumber
  &\leq& \mathop{\sup_{U \leftarrow A}}_{V \leftarrow U} H(U|VFT) -
  H(U|VBT) \\
  \label{public_announcement}
  &\leq& \mathop{\sup_{U \leftarrow AT}}_{V \leftarrow U} H(U|VFT) - H(U|VBT).
\end{eqnarray}
In the first line we just use the definition of the classical secret
key rate given by Eq.~(\ref{rate_classicalbound}). The first
inequality comes from the fact that conditioning can only decrease the
entropy, \textit{i.e.}, $H(U|VB) \geq H(U|VBT)$. For the last
inequality, we give Alice also access to the random variable $T$,
additionally to her variable $A$, over which she can perform the
post-processing. Altogether Eq.~(\ref{public_announcement}) tells that
if Eve announces publicly the value of the variable $T$, so whether
Alice and Bob share the extendible or non-extendible part, that this
action can only enhance Alice and Bob's ability to create a secret key.

Next, we have that 
\begin{eqnarray}
  \nonumber
  &&\mathop{\sup_{U \leftarrow AT}}_{V \leftarrow U} H(U|VFT) - H(U|VBT) \\
  \nonumber
  &=& \mathop{\sup_{U \leftarrow AT}}_{V \leftarrow U} \sum p(t_k)
  \big\{ H(U|VFt_k) - H(U|VBt_k) \big\} \\
  \nonumber
  &=& \sum p(t_k) \mathop{\sup_{U \leftarrow At_k}}_{V \leftarrow
    U} \big\{ H(U|VFt_k) - H(U|VBt_k) \big\} \\
  \label{average}
  &=& \sum p(t_k) S^{t_k}_\rightarrow(A;B|F).
\end{eqnarray}
First we rewrite the conditional entropies in terms of an
expectation value over the parameter $t_k$. The map $U \leftarrow AT$
acts independent on each term of the sum over $t_k$. Therefore the
supremum can be put into the sum taking the specific value of
$t_k$. Since $\sup_{U \leftarrow A t_k}$ is equal to $\sup_{U
  \leftarrow A}$ for $t_k$ fixed, we find on the right hand side the
one-way secret key rate for the conditional three party correlation
$p(a_i,b_j,f_l|t_k)$ denoted as $S^{t_k}_\rightarrow(A;B|F)$.

Combining Eqs.~(\ref{public_announcement},\ref{average}) we
have, therefore, that
\begin{equation}
  S_\rightarrow(A;B|FT) \leq \sum p(t_k) S_\rightarrow(A;B|F;T=t_k).
\end{equation}
From Observation $1$ we learn that Alice and Bob cannot draw a secret
key out of the extendible part $\sigma_{ext}$, \textit{i.e.},
$S_\rightarrow(A;B|E;t_k=ext)=0$. Therefore, only the non-extendible
part $\rho_{ne}$ can contribute to a positive secret key rate. The
conditional probability distribution
$p(a_i,b_j,f_k|t_k=ne)\equiv\tilde p_{ijk}$ defines exactly the
considered extensions $\mathcal{P}^*$. This concludes the
proof. $\blacksquare$  

The upper bound given by Eq. (\ref{bound_theorem1}) requires to
solve the infimum over all possible extensions
$\mathcal{P^*}$. Instead of this optimization, one can just pick a
particular state $\mathcal{S}_{max}^{ne}$ and calculate the infimum
over all possible measurements $\{E_k\}$ employed by Eve.  
 
\textit{Corollary 1}: Given a state $\rho_{ne} \in \mathcal{S}_{max}^{ne}$,
the secret key rate $K_{\rightarrow}$ is bounded from above by
\begin{equation}
  \label{bound_corollary1} 
  K_{\rightarrow} \leq (1-\lambda_{max}^{\mathcal{S}}) \inf_{E_k}
  S_{\rightarrow}^{E_k}(A;B|E),  
\end{equation} 
with $S_{\rightarrow}^{E_k}(A;B|E)$ being the classical one-way secret key
rate of the tripartite probability distribution 
$\tilde p_{ijk}=Tr( A_i\otimes B_j\otimes E_k\ 
(\ket{\Phi_{ne}}\bra{\Phi_{ne}}))$, and where $\ket{\Phi_{ne}}$ denotes a
purification of $\rho_{ne}$.  
 
\textit{Proof}:  The right hand side of Eq. (\ref{bound_corollary1})
is an upper bound of the right hand side of
Eq. (\ref{bound_theorem1}), because in Eq. (\ref{bound_corollary1})
we take only a particular state $\rho_{ne} \in
\mathcal{S}_{max}^{ne}$, whereas in Eq. (\ref{bound_theorem1}) we
perform the infimum over all possible states $\rho_{ne}
\in \mathcal{S}_{max}^{ne}$. $\blacksquare$  
 
The upper bounds provided by Theorem $1$ and Corollary $1$ still
demand solving a difficult optimization problem. Next, we present a
simple upper bound on $K_{\rightarrow}$ that is straightforward to
calculate. Then, in Sec.~\ref{section_evaluation}, we illustrate the
performance of this upper bound for two well-known QKD protocols: the
four-state \cite{bennett84a} and the six-state \cite{bruss98a} QKD
schemes. We compare our results to other well-known upper bounds on
$K_{\rightarrow}$ for these two QKD schemes
\cite{fuchs97a,cirac97a,bech99a,kraus04a,renner05a}.   
 
\textit{Corollary 2}: The secret key rate $K_\rightarrow$ is upper bounded by 
\begin{equation}
  \label{bound_corollary2}
  K_\rightarrow \leq (1-\lambda_{max}^{\mathcal{S}})\ I^{ne}(A;B), 
\end{equation} 
where $I^{ne}(A;B)$ denotes the classical mutual information calculated on the 
probability distribution $\tilde p_{ij}=Tr( A_i \otimes B_j\ \rho_{ne})$
with $\rho_{ne} \in \mathcal{S}_{max}^{ne}$.  
 
\textit{Proof}: According to Eq. (\ref{rate_mutualinformation}), the one-way
secret key rate $S_\rightarrow(A;B|E)$ is bounded from above by the mutual
information $I(A;B)$. Note that the mutual information $I(A;B)$ is an upper
bound on the secret key rate for \emph{arbitrary} communication protocols
\cite{maurer99a}. $\blacksquare$ 
 
The main difficulty when evaluating the upper bound given by 
Eq.~(\ref{bound_corollary2}) 
for a particular
realization of QKD relies on obtaining the parameter
$\lambda_{max}^{\mathcal{S}}$ and the non-extendible state $\rho_{ne}$. In
App. \ref{ap_sdpmethod} we show how this problem can be cast as a convex
optimization problem known as semidefinite program \footnote{In order to 
verify only whether the parameter
  $\lambda_{max}^{\mathcal{S}}$ is zero or not, one can use as well
  entanglement witnesses of a particular form
  \cite{moroder05_thesis}.}. Such instances of convex optimization problems
can be efficiently solved, for example by means of interior-point methods  
\cite{vandenberghe96a, vandenberghe04a}.

\section{Evaluation}
\label{section_evaluation} 
 
In this section we evaluate the upper bound on $K_\rightarrow$ given by
Eq. (\ref{bound_corollary2}) for two well-known qubit-based QKD protocols: the
four-state \cite{bennett84a} and the six-state \cite{bruss98a} QKD schemes. 
We select these two particular QKD schemes because they allow us to compare
our results with already known upper bounds on $K_{\rightarrow}$
\cite{fuchs97a,cirac97a,bech99a,kraus04a,renner05a}. Let us emphasize, 
however,
that our method can also be used straightforwardly to obtain an upper bound for
higher dimensional, more complicated QKD protocols, for which no upper bounds
have been calculated yet. By means of semidefinite programming one can 
easily obtain
the maximum weight of extendibility $\lambda^{\cal S}_{max}$ and the
corresponding non-extendible part $\rho_{ne}$ which suffice for the computation
of the upper bound. (See App.~\ref{ap_sdpmethod}.) 

In the case of the four-state EB protocol, Alice and Bob perform projection
measurements onto two mutually unbiased bases, say the ones given by the
eigenvectors of the two Pauli operators $\sigma_x$ and $\sigma_z$. In the
corresponding PM scheme, Alice can use as well the same set of measurements
but now on a maximally entangled state. Note that here we are not
using the general approach introduced previously,
$\ket{\psi_{source}}_{AB}=\sum_i \sqrt{p_i}
\ket{\alpha_i}_A\ket{\varphi_i}_{B}$, to model PM schemes, since for
these two protocols it is sufficient to consider that the effectively
distributed quantum states consist only of two qubits.  For the case
of the six-state EB protocol, Alice and Bob perform projection
measurements onto the eigenvectors of the three Pauli operators
$\sigma_x, \sigma_y,$ and $\sigma_z$ on the bipartite qubit states
distributed by Eve. In the corresponding PM scheme 
Alice prepares the eigenvectors of those operators by performing the same
measurements on a maximally entangled two-qubit state. 
 
We model the transmission channel as a depolarizing channel with error
probability $e$. This defines one possible eavesdropping interaction. In
particular, the observed probability distribution $p_{ij}$ is obtained
in both protocols by measuring the quantum state $\rho_{AB}(e)=(1-2e)
\ket{\psi^+}_{AB}\bra{\psi^+}+e/2 \openone_{AB}$, where the state
$\ket{\psi^+}_{AB}$ represents a maximally entangled two-qubit state as
$\ket{\psi^+}_{AB}=1/\sqrt{2}(\ket{00}_{AB}+\ket{11}_{AB})$, 
and $\openone_{AB}$ denotes the identity operator. The state
$\rho_{AB}(e)$ provides a probability distribution $p_{ij}$ that is invariant
under interchanging Alice and Bob. This means that for this particular example
there is no difference whether we consider direct communication (extension of
$\rho_{AB}(e)$ to two copies of system $B$) or reverse reconciliation
(extension of $\rho_{AB}(e)$ to two copies of system $A$). The quantum bit
error rate ($QBER$), \textit{i.e.}, the fraction of signals where Alice and 
Bob's measurements results differ, is given by $QBER=e$. 

\begin{figure} 
  \centering 
  \includegraphics[scale=0.7]{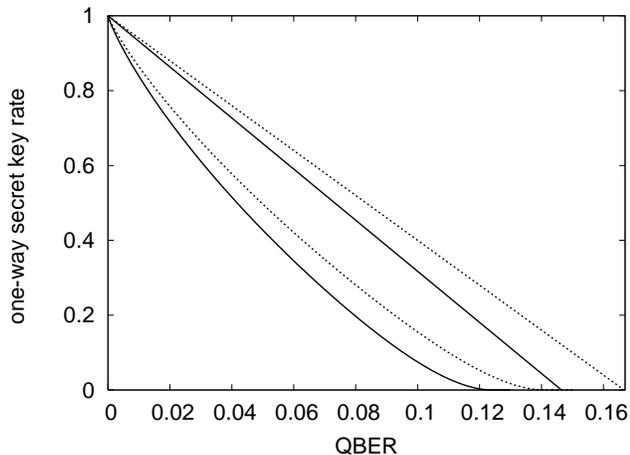} 
  \caption{Upper bound on the one-way secret key rate $K_\rightarrow$ given by 
  Eq.~(\ref{bound_corollary2}) for the four-state (solid) and the six-state
  (dotted) QKD protocols in comparison to known lower bounds on the
  secret key rate given in Ref.~\cite{renner05a}. The equivalence
  class of states $\mathcal{S}$ is fixed by the observed data
  $p_{ij}$, which are generated via measurements onto the state $\rho_{AB}(e) =
  (1-2e)\ket{\psi^+}_{AB}\bra{\psi^+} + e/2 \openone_{AB}$. The quantum bit
  error rate is given by $QBER=e$. Here we assume an asymmetric basis
  choice to suppress the sifting effect \cite{lo05a}.}  
  \label{figure_example} 
\end{figure} 
  
The resulting upper bounds on $K_\rightarrow$ are illustrated in 
Fig. \ref{figure_example}. These results do not include the sifting factor of
$1/2$ for the four-state protocol ($1/3$ for the six-state protocol), since
this effect can be avoided by an asymmetric basis choice for Alice and Bob
\cite{lo05a}.  

Let us consider in more detail the cut-off points for $K_\rightarrow$,
\textit{i.e.}, the values of $QBER$ for which the secret key rate drops down
to zero in Fig.~\ref{figure_example}. We find that in the four-state protocol
(six-state protocol) one-way secret key distillation might only be possible
for a $QBER<14.6$ ($QBER<1/6$). These results reproduce the well-known upper
bounds on both protocols from
Refs.~\cite{fuchs97a,cirac97a,bech99a}. More recently, a new upper
bound for the six-state protocol was obtained in
Refs.~\cite{kraus04a,renner05a}, $QBER \lessapprox 16.3$. This result
indicates that the upper bound given by Eq. (\ref{bound_corollary2})
is not tight, since it fails to reproduce this last value. More
importantly, both statements together entail that Observation $1$ is
necessary but not sufficient for one-way secret key distillation:
there exist bipartite states which are non-extendible, nevertheless no
secret key can be obtained from them via one-way post-processing. It
shows that the characterization of useful quantum states for one-way
QKD is still an open problem. Finally, let us mention that the
threshold point for the four-state protocol computed in
Ref.~\cite{kraus04a,renner05a} lead to the same cut-off point as for
the optimal approximate cloner attack and thus also coincides with the
value of our result.

\section{Conclusion}
\label{section_conclusion}
 
In this paper we address the fundamental question of how much secret
key can be obtained from the classical data that become available once
the first phase of QKD is completed. In particular, we restrict
ourselves to one-way public communication protocols between the
legitimate users. This question has been extensively studied in the
literature and analytic expressions for upper and lower bounds on the
one-way secret key rate are already known. Unfortunately, to evaluate
these expressions for particular QKD protocols is, in general, a
non-trivial task. It demands to solve difficult optimization problems
for which no general solution is known so far.  

Here we provide a simple method to obtain an upper bound on the
one-way secret key rate for QKD. It is based on a necessary
precondition for one-way secret key distillation: The legitimate users
need to prove that there exists no quantum state having a symmetric
extension that is compatible with the available measurements
results. The main advantage of the method is that it is
straightforward to calculate, since it can be formulated as a
semidefinite program. Such instances of convex optimization problems
can be solved very efficiently. More importantly, the method applies
both to prepare and measure schemes and to entanglement based schemes,
and it can reproduce most of the already known cut-off points for
particular QKD protocols. Recent results show that the given
precondition is only necessary but not sufficient, so there exists
non-extendible quantum state which nevertheless are useless for
one-way key distillation.

\section{Acknowledgements} 
 
The authors wish to thank R. Renner, A. Winter, M. Koashi, J.~I. Cirac,
D. Bru{\ss}, J.~M. Renes, W. Mauerer, J. Rigas, H. H\"aseler and V. Scholz for
very useful discussions, and especially G. O. Myhr for very interesting 
discussions about App.~\ref{ap_mixedstates}.
This work was  supported by the DFG under the Emmy
Noether programme, and the  European Commission (Integrated Project SECOQC).

\appendix 

\section{QKD with mixed signal states}\label{ap_mixedstates}

In this Appendix we describe very briefly the case of QKD 
based on mixed quantum states instead of pure states. In particular, 
we analyze PM schemes, since in EB 
schemes it is clear that Eve can always distribute mixed states 
to Alice and Bob, and this situation is already contained in the 
results included in the previous sections. More specifically, we 
translate the necessary precondition for secret key generation by
unidirectional communication to the PM mixed state scenario.   

In the most general PM scheme, Alice prepares mixed signal states $\rho_B^i$
following a given probability distribution $p_i$ and sends them to 
Bob. Equivalently, we can think of the preparation process as follows. Suppose
that the spectral decomposition of the signal state $\rho_B^i$ is given by
$\rho^i_B=\sum_j \lambda_j^i \ket{\varphi_j^i}_B\bra{\varphi_j^i}$. This 
can be interpreted as producing with probability $\lambda_j^i$ the pure state
$\ket{\varphi_j^i}$. Alternatively, $\rho_B^i$ can as well
originate from a pure state in a higher dimensional Hilbert space. 
That is, from a possible purification $\ket{\phi_i}_{A^\prime B}$ of the
state $\rho_B^i$ in the composite Hilbert space $\mathcal{H}_{A^\prime}
\otimes \mathcal{H}_B$ which reads as 
\begin{equation}
  \ket{\phi^i}_{A^\prime B}=\sum_j \sqrt{\lambda_j^i} \ket{j}_{A^\prime}
  \ket{\varphi_j^i}_B.  
\end{equation}
Now we can use the same formalism as the one for PM schemes based on pure
signal states. 
We can assume that first Alice prepares the tripartite quantum state   
\begin{equation}
  \ket{\psi_{source}}_{AA^\prime B}=\sum_{ij} \sqrt{p_i \lambda_j^i} \ket{i}_A
  \ket{j}_{A^\prime} \ket{\varphi_j^i}_B.
\end{equation}
Afterwards, in order to produce the actual signal state in system $B$, 
Alice performs a
measurement onto the standard basis of system $A$ only, and completely ignores 
system $A^\prime$. Her measurement operators are given by $A_i=\ket{i}_A\bra{i}
\otimes \openone_{A^\prime}$. The produced signal states are sent to Bob who
measures them by means of the POVM $\{ B_j \}$. Since Eve can only
interact with system $B$, the reduced density matrix of $\rho_{AA^\prime}
= Tr_B(\ket{\psi_{source}}_{AA^\prime B} \bra{\psi_{source}})$ is fixed by the
actual preparation scheme. This information can be included in the measurement
process by adding to the observables measured by Alice and Bob other
observables $\{ C_{k,AA^\prime} \otimes \openone_B \}$ such that they provide
complete information on the bipartite Hilbert space of Alice
$\mathcal{H}_{AA^\prime}=\mathcal{H}_{A}\otimes\mathcal{H}_{A^\prime}$. 
(See also \footnote{Note that the preparation scheme is not
unique. There are many different ways to actually purify all the
signal states $\rho_B^i$ together. For example every signal state can
be purified to the same auxiliary system $A^\prime$, alternatively
each signal can as well be purified to a special separated system, say
$A^\prime_i$, so that the overall purified system $A^\prime=A^\prime_1 \cdots
A^\prime_n$ is higher dimensional. This will lead to different
reduced density matrices $\rho_{AA^\prime}$. Note however that the
exact form of reduced density matrix $\rho_{AA^\prime}$ has no
physical meaning, except for the a priori probabilities of the signals
on the diagonal.}.)

The relevant equivalence class of quantum states 
$\mathcal{S}_{AA^\prime B}$ contains all
tripartite quantum states $\rho_{AA^\prime B}$ consistent with the available
information during the measurement process. Obviously, Eve can 
always access every
purification $\ket{\Psi_{AA^\prime BE}}$ of a state in
$\mathcal{S}_{AA^\prime B}$. Note that the situation is 
completely equivalent to the
case of pure signal states \cite{curty04a}.  

Now we are ready to rephrase the necessary precondition for one-way secret key
distillation for the case of QKD based on mixed states. 
For direct communication we need to search for symmetric extensions to
two copies of system $B$. That is, if we denote with $\bar A$ the
bipartite system on Alice's side $\bar A \equiv AA^\prime$, we have to
search for quantum states in the equivalence class
$\mathcal{S}_{AA^\prime B}=\mathcal{S}_{\bar A B}$ which are
extendible to $\rho_{\bar A B B^\prime}$. In the case of reverse
reconciliation, Eve needs to possess only a copy of system $A$. Note
that the final key is created only from measurements onto this
system. Therefore, in order to determine the cut-off points for the
key distillation process, one has to examine the question whether a
four-partite quantum state $\rho_{AA^\prime BE}$ with
$Tr_E(\rho_{AA^\prime BE}) \in \mathcal{S}_{AA^\prime B}$ exists such
that $Tr_{A^\prime}(\rho_{AA^\prime BE})$ is exactly the desired
symmetric extension to two copies of system $A$.

\section{Semidefinite Programs And Searching For
  $\lambda^{\mathcal{S}}_{max}$  And $\rho_{ne}$}\label{ap_sdpmethod}  
 
In this Appendix we provide a method to obtain the parameter
$\lambda^{\mathcal{S}}_{max}$ and the corresponding non-extendible state
$\rho_{ne}$. In particular, we show how one can cast this problem as a
convex optimization problem known as semidefinite programming. Such instances
of convex optimization problems appear frequently in quantum information
theory and they can be solved very efficiently. There are freely-available
input tools like, for instance, YALMIP \cite{yalmip}, and standard
semidefinite programming solvers, see SeDuMi \cite{sedumi} and SDPT3-3.02
\cite{sdpt}. 
 
\newbox\schnuffibox
\setbox\schnuffibox=\hbox{\begin{minipage}{\columnwidth-0.7cm}\footnotesize\begin{equation*}F({\bf{x}})=\left( \begin{array}{c|c}   
\tilde F({\bf{x}}) & 0  \\ \hline 
  0 & \bar F({\bf{x}})  \end{array} \right) \equiv \tilde F({\bf{x}}) \oplus
  \bar F({\bf{x}}) \geq 0.
\end{equation*}\end{minipage}}
\noindent{}

A typical semidefinite problem, also known as primal problem, has the
following form  
\begin{eqnarray} 
  \label{primalSDP} 
  \text{minimise} && c^T {\bf{x}} \\ 
  \nonumber \text{subject to} && F({\bf{x}})=F_0 + \sum_i x_i F_i \geq 0, 
\end{eqnarray} 
where the vector ${\bf x}=(x_1, ..., x_t)^T$ represents the objective
variable, the vector $c$ is fixed by the particular optimization
problem, and the matrices $F_0$ and $F_i$ are Hermitian matrices. The
goal is to minimize the linear objective function $c^T{\bf{x}}$
subject to a linear matrix inequality constraint $F({\bf{x}})
\geq 0$ \cite{vandenberghe96a, vandenberghe04a}. (See also 
Ref. \footnote{Note that two (or even more) linear matrix inequalities
  constraints $\tilde F({\bf{x}}) \geq 0, \bar F({\bf{x}}) \geq 0$,
  can be combined into a single new linear matrix inequality
  constraint as: \usebox\schnuffibox}.) 

Any bounded Hermitian operator $\rho_A=\rho_A^\dag$ acting on a
$n$-dimensional Hilbert space $\mathcal{S}$ can be written in terms of an
operator basis, which we shall denote by $\{S_k\}$, satisfying the
following three conditions: $Tr(S_j)=n \delta_{j1}$, $S_j=S_j^\dag$,
and $Tr(S_j S_{j^\prime})=n \delta_{jj^\prime}$. A possible choice is
given by the $SU(n)$ generators. Using this representation, a general
bipartite state $\rho_{AB}$ in a $d_{AB}$-dimensional Hilbert space
can be written as    
\begin{equation}\label{eqnew}
  \rho_{AB}=\frac{1}{d_{AB}} \sum_{kl} r_{kl} S^A_k S^B_l,
\end{equation}
where the coefficients $r_{kl}$ are given by $r_{kl}=Tr(S^A_k S^B_l
\rho_{AB})$. Note that in order to simplicity the notation, in this 
Appendix we omit the tensor product signs $\otimes$ between the 
operators. The representation given by Eq. (\ref{eqnew}) allows us to describe
any bipartite density operator in terms of a fixed number of parameters
$r_{kl}$ which can serve as the free parameters in the program.

The knowledge of Alice and Bob's POVMs $\{A_i\}$ and $\{ B_j\}$, respectively, 
and the observed probability distribution $p_{ij}$ determines the equivalence
class of compatible states $\mathcal{S}$. Since every POVM element is a
Hermitian operator itself, every element can as well be expanded in the
appropriate operator basis $A_i = \sum_m a_{im} S_m^A$ and $B_j = \sum_n b_{jn}
S_n^B$. 

An arbitrary operator $\rho_{AB}=1/d_{AB} \sum r_{kl} S_k^A S_j^B$
belongs to the equivalence class $\mathcal{S}$ if and only if it fulfils
the following constraints: In order 
to guarantee that the operator $\rho_{AB}$ represents a valid 
quantum state, it must be normalized $Tr(\rho_{AB})=r_{11}=1$, and it must be a
semidefinite positive operator $\rho_{AB} \geq 0$. In addition, it must
reproduce the observed data of Alice and Bob. This last 
requirement imposes the following
constraints  
\begin{equation} 
  Pr(a_i,b_j)= \sum_{kl} a_{ik} b_{il} r_{kl} = p_{ij}\ \ \forall i,j,
\end{equation} 
which are linear on the state coefficients $r_{kl}$. Note that every equality
constraint $Pr(a_i,b_j)=p_{ij}$
can be represented by two inequality constraints as
$Pr(a_i,b_j)-p_{ij}\geq 0$ and $-(Pr(a_i,b_j)-p_{ij})\geq 0$. 
 
In order to find the decomposition of a given state $\rho_{AB}=1/d_{AB}
\sum_{kl} r_{kl} S_k^A S_l^B$ into an extendible part $\sigma_{ext}$ and an
non-extendible part $\rho_{ne}$, with maximum weight $\lambda_{max}(\rho_{AB})$
of extendibility, we can proceed as follows. First we rewrite the
problem in terms of unnormalized states $\tilde \sigma_{ext} \equiv
\lambda \sigma_{ext}$ and $\tilde \rho_{ne} \equiv (1-\lambda)\rho_{ne}$ as
\begin{equation} 
  \label{bse_short}
  \rho_{AB} = \min_{Tr(\tilde \rho_{ne})} \tilde \sigma_{ext} + \tilde \rho_{ne}. 
\end{equation} 
Now assume that the unnormalized, extendible state is written as $\tilde
\sigma_{ext}=1/d_{AB} \sum \tilde e_{kl} S_k^A S_l^B$, which must form a
semidefinite positive operator $\tilde \sigma_{ext} \geq 0$. In the case of
direct communication we have to impose that $\tilde \sigma_{ext}$ has a
symmetric extension $\chi_{ABB^\prime}$ to two copies of system B. That is,
$\chi_{ABB^\prime}$ remains invariant under permutation of the systems $B$ and
$B^\prime$. This is only possible if the state $\chi_{ABB^\prime}$ can
be written as  
\begin{eqnarray}
  \nonumber
  \chi_{ABB^\prime}&=&\frac{1}{d_{ABB^\prime}} \sum_{\substack{k \\ l>m}}
  f_{klm} (S_k^A S_l^B S_m^{B^\prime} + S_k^A S_m^B S_l^{B^\prime}) \\ 
  &+& \sum_{kl} f_{kll} S_k^A S_l^B S_l^{B^\prime}
\end{eqnarray}
with appropriate state coefficients $f_{klm}\ \forall k, \forall l \geq m$. The
extension must as well reproduce the original state $Tr_{B^\prime}
(\chi_{ABB^\prime}) = \tilde \sigma_{ext}$, which implies that the state
coefficients of $\tilde \sigma_{ext}$ and $\chi_{ABB^\prime}$ are related by 
\begin{equation}
  f_{kl1}=\tilde e_{kl}\ \forall k,l.
\end{equation}
Hence, some of the state parameters of $\chi_{ABB^\prime}$ are already fixed by
the coefficients of $\tilde \sigma_{ext}$. In addition, the coefficients
$f_{klm}$ have to form a semidefinite positive operator $\chi_{ABB^\prime}
\geq 0$.

Once the states $\rho_{AB}=\sum r_{kl} S^A_k S^B_l$ and the 
unnormalized extendible part $\tilde \sigma_{ext}=\sum e_{kl} S^A_k 
S^B_l$ are fixed, the remaining non-extendible state $\tilde
\rho_{ne}$ is determined by the decomposition given by
Eq. (\ref{bse_short}), and equals to  
\begin{equation} 
  \tilde \rho_{ne}=\rho_{AB}-\sum (r_{kl} - e_{kl}) S^A_k S^B_l. 
\end{equation} 
This operator must be semidefinite positive $\rho_{ne} \geq 0$. 

In total, the state coefficients of the states in the equivalence class
$\rho_{AB}$, the unnormalized, extendible part in the best extendibility
decomposition $\tilde \sigma_{ext}$ and the symmetric extension itself
$\chi_{ABB^\prime}$ will constitute the objective variables of the SDP
program
\begin{equation}
  {\bf{x}} = ( r_{kl}, \tilde e_{kl}, f_{klm})^T.
\end{equation}
This requires a fixed ordering of the set of coefficients. The function to be
minimized is the weight on the unnormalized, non-extendible part, $Tr(\tilde
\rho_{ne})=r_{11}-\tilde e_{11}$. Hence the vector $c$ is given by 
\begin{equation}
  c^T=( \underbrace{1}_{r_{11}},0, \cdots, \underbrace{-1}_{\tilde e_{11}}, 0.
  \cdots ),
\end{equation}
All the semidefinite constraints introduced previously on the state
coefficients can be collected into a single linear matrix inequality
constraint given by  Eq. (\ref{primalSDP}). The final $F({\bf{x}})$
collects all these constraints as block-matrices on the diagonal.

\bibliographystyle{apsrev}

\end{document}